\documentstyle[12pt]{article}
\newcommand{\Cslash}{\not \!\! C}
\newcommand{\Dslash}{\not \!\! D}
\newcommand{\kslash}{\not \!\! k}
\newcommand{\sslash}{\not \!\! s}
\begin{document}

\begin{flushright}{UT-889\\
May 2000
}
\end{flushright}

\begin{center}
{\large{\bf Chiral Anomaly for a New Class of 
Lattice Dirac Operators }}
\end{center}
\vskip .5 truecm
\centerline{\bf Kazuo Fujikawa and Masato Ishibashi}
\vskip .4 truecm
\centerline {\it Department of Physics,University of Tokyo}
\centerline {\it Bunkyo-ku,Tokyo 113,Japan}
\vskip 0.5 truecm

\makeatletter
\@addtoreset{equation}{section}
\def\theequation{\thesection.\arabic{equation}}
\makeatother

\begin{abstract}
A new class of lattice Dirac operators which satisfy the index 
theorem  have been recently proposed on the basis of the 
algebraic relation
$\gamma_{5}(\gamma_{5}D) + (\gamma_{5}D)\gamma_{5} =
2a^{2k+1}(\gamma_{5}D)^{2k+2}$. Here $k$ stands for a 
non-negative integer and $k=0$ corresponds to the ordinary
Ginsparg-Wilson relation. We analyze the chiral anomaly and 
 index theorem for all these Dirac operators in an explicit 
elementary manner. We show that the coefficient of anomaly is 
 independent of a small variation in the parameters $r$ and 
$m_{0}$, which characterize these Dirac operators, and the 
correct chiral anomaly is obtained in the (naive) continuum limit 
$a\rightarrow 0$.
\end{abstract}

\section{Introduction}
A new class of lattice Dirac operators $D$ have been
recently proposed on the basis of the algebraic relation[1]
\begin{equation}
\gamma_{5}(\gamma_{5}D)+(\gamma_{5}D)\gamma_{5}=2a^{2k+1}
(\gamma_{5}D)^{2k+2}
\end{equation}
where $k$ stands for a non-negative integer, and $k=0$ corresponds
to the ordinary Ginsparg-Wilson relation[2] for which an explicit 
example of the operator free of species doubling has been given 
by Neuberger[3]. It has been shown in [1] that we can in fact 
construct the lattice Dirac operator, which is free of species 
doublers, for all values of $k$. Here  $\gamma_{5}$ is a 
hermitian chiral Dirac matrix and $\gamma_{5}D$ is also 
hermitian. 

When one defines 
\begin{equation}
\Gamma_{5}\equiv\gamma_{5}-(a\gamma_{5}D)^{2k+1}
\end{equation}
the relation (1.1) is written as 
\begin{equation}
\Gamma_{5}(\gamma_{5}D)+(\gamma_{5}D)\Gamma_{5}=0.
\end{equation}
The index relation[4][5] on the lattice is generally written as
\begin{equation}
Tr\Gamma_{5}=n_{+}-n_{-},
\end{equation}
which is confirmed by[1] 
\begin{eqnarray}
Tr\Gamma_{5}&\equiv&
\sum_{\lambda_{n}}\phi^{\dagger}_{n}\Gamma_{5}\phi_{n}
\nonumber\\
&=&\sum_{ \lambda_{n}=0}\phi^{\dagger}_{n}\Gamma_{5}\phi_{n}+
\sum_{\lambda_{n}\neq 0}\phi^{\dagger}_{n}\Gamma_{5}\phi_{n}
\nonumber\\
&=&\sum_{\lambda_{n}=0}\phi^{\dagger}_{n}\Gamma_{5}\phi_{n}
\nonumber\\
&=&\sum_{\lambda_{n}=0}\phi^{\dagger}_{n}[\gamma_{5}
-(a\gamma_{5}D)^{2k+1}]\phi_{n}\nonumber\\
&=&n_{+}-n_{-}=index
\end{eqnarray}
where $n_{\pm}$ stand for the number of  normalizable zero modes 
in 
\begin{equation}
\gamma_{5}D\phi_{n}=0
\end{equation}
for the {\em hermitian}  operator $\gamma_{5}D$ with 
simultaneous eigenvalues $\gamma_{5}\phi_{n}= \pm \phi_{n}$.  
We also used the relation following from (1.3) 
\begin{equation}
\gamma_{5}D\Gamma_{5}\phi_{n}=-\lambda_{n}\Gamma_{5}\phi_{n}
\end{equation}
if 
\begin{equation}
\gamma_{5}D\phi_{n}=\lambda_{n}\phi_{n},
\end{equation}
which suggests that either $\Gamma_{5}\phi_{n}$ for 
$\lambda_{n}\neq 0$ is orthogonal to $\phi_{n}$ or else 
$\Gamma_{5}\phi_{n}=0$.
The positive definite inner product is defined by summing over 
all the lattice points
\begin{equation}
\phi^{\dagger}_{n}\phi_{n}=(\phi_{n},\phi_{n})
\equiv\sum_{x}a^{4}\phi^{\star}_{n}(x)\phi_{n}(x)
\end{equation}
but the coordinate $x$ is often omitted in writing $\phi_{n}$. 

The Euclidean path integral for a fermion is defined by
\begin{equation}
\int{\cal D}\bar{\psi}{\cal D}\psi\exp[\int\bar{\psi}D\psi]
\end{equation}
where
\begin{equation}
\int\bar{\psi}D\psi\equiv \sum_{x,y}\bar{\psi}(x)D(x,y)\psi(y)
\end{equation}
and the summation runs over all the points on the lattice.
The relation (1.3) is re-written as 
\begin{equation}
\gamma_{5}\Gamma_{5}\gamma_{5}D+D\Gamma_{5}=0
\end{equation}
and thus the Euclidean action is invariant under the global
 ``chiral'' transformation[5]
\begin{eqnarray}
&&\bar{\psi}(x)\rightarrow\bar{\psi}^{\prime}(x)=
\bar{\psi}(x)+i\sum_{z}\bar{\psi}(z)\epsilon\gamma_{5}
\Gamma_{5}(z,x)\gamma_{5}
\nonumber\\
&&\psi(y)\rightarrow\psi^{\prime}(y)=
\psi(y)+i\sum_{w}\epsilon\Gamma_{5}(y,w)\psi(w)
\end{eqnarray}
with an infinitesimal constant parameter $\epsilon$.
Under this transformation, one obtains a Jacobian factor
\begin{equation}
{\cal D}\bar{\psi}^{\prime}{\cal D}\psi^{\prime}=
J{\cal D}\bar{\psi}{\cal D}\psi
\end{equation}
with
\begin{equation}
J=\exp[-2iTr\epsilon\Gamma_{5}]=\exp[-2i\epsilon(n_{+}-n_{-})]
\end{equation}
where we used the index relation (1.5). This derivation may be 
regarded as a lattice counter part of the continuum path 
integral[6].

In Ref.[1] it was shown by using the method in [7], which is 
a lattice extension of the method in [6], that 
the index $n_{+}-n_{-}$ appearing in the Jacobian factor is 
related to the Pontryagin number for any
 operator in (1.1) if the operator $\gamma_{5}D$ satisfies 
suitable conditions.  In this paper, we evaluate $Tr \Gamma_{5}$ 
in a more explicit and elementary manner on the basis of explicit 
formulas for $\gamma_{5}D$ in the continuum 
limit\footnote{The continuum limit in this paper stands for the 
so-called ``naive''continuum limit with $a\rightarrow 0$, and 
the lattice size is gradually extended to infinity for any 
finite $a$ in the process of taking the limit $a\rightarrow 0$.}. 
We show that these operators $\gamma_{5}D$ for all $k$ in fact 
reproduce the correct chiral anomaly and consequently correct 
Pontryagin number.

\section{A brief summary of the model and notation}

The operator $\Gamma_{5}$ appearing in the index relation (1.5)
has an explicit expression[1]
\begin{equation}
\Gamma_{5}=\gamma_{5}-H_{(2k+1)}
\end{equation}
with
\begin{equation}
H_{(2k+1)}\equiv(\gamma_{5}aD)^{2k+1}=\frac{1}{2}\gamma_{5}
[1+D_{W}^{(2k+1)}\frac{1}
{\sqrt{(D_{W}^{(2k+1)})^{\dagger}D_{W}^{(2k+1)}}}].
\end{equation}
The operator $D_{W}^{(2k+1)}$ is in turn expressed as a
generalization of the ordinary Wilson Dirac operator
as 
\begin{equation}
D_{W}^{(2k+1)}=i(\Cslash)^{2k+1}+(B)^{2k+1}
-(\frac{m_{0}}{a})^{2k+1}.
\end{equation}
See Appendix for further details of the general solution to 
(1.1).  

The ordinary Wilson Dirac operator $D_{W}$, which corresponds to 
$D_{W}^{(1)}$, is given by
\begin{eqnarray}
D_{W}(x,y)&\equiv&i\gamma^{\mu}C_{\mu}(x,y)+B(x,y)-
\frac{1}{a}m_{0}\delta_{x,y},\nonumber\\
C_{\mu}(x,y)&=&\frac{1}{2a}[\delta_{x+\hat{\mu} a,y}
U_{\mu}
(y)-\delta_{x,y+\hat{\mu} a}U^{\dagger}_{\mu}(x)],
\nonumber\\
B(x,y)&=&\frac{r}{2a}\sum_{\mu}[2\delta_{x,y}-
\delta_{y+\hat{\mu} a,x}U_{\mu}^{\dagger}(x)
-\delta_{y,x+\hat{\mu} a}U_{\mu}(y)],
\nonumber\\
U_{\mu}(y)&=& \exp [iagA_{\mu}(y)],
\end{eqnarray}
where we added a constant mass term to $D_{W}$.  Our 
matrix convention is that $\gamma^{\mu}$ are anti-hermitian, 
$(\gamma^{\mu})^{\dagger} = - \gamma^{\mu}$, and thus 
$\Cslash\equiv \gamma^{\mu}C_{\mu}(n,m)$ is hermitian
\begin{equation}
\Cslash ^{\dagger} = \Cslash.
\end{equation}
Since the operators $\Cslash$ and $B$ form the basis for any 
fermion operator on the lattice, we summarize the basic 
properties of $\Cslash$ and $B$.

\subsection{Operators  $\Cslash$ and $B$ and  Brillouin zone}

For a square lattice, for which we work in this paper, one can 
explicitly show that the simplest lattice fermion action
\begin{equation}
S =\int\bar{\psi}i\Cslash\psi
\end{equation}
is invariant under the transformation[8]
\begin{equation}
\psi^{\prime}= {\cal T}\psi,\ \bar{\psi}^{\prime}= \bar{\psi}{\cal T}^{-1}
\end{equation}
where ${\cal  T}$ stands for any one of the following 16 operators
\begin{equation}
1,\ T_{1}T_{2},\ T_{1}T_{3},\ T_{1}T_{4},\ T_{2}T_{3},
\ T_{2}T_{4},\ T_{3}T_{4},\ T_{1}T_{2}T_{3}T_{4},
\end{equation}
and 
\begin{equation}
T_{1},\ T_{2},\ T_{3},\ T_{4},\ T_{1}T_{2}T_{3},
\ T_{2}T_{3}T_{4},\ T_{3}T_{4}T_{1},\ T_{4}T_{1}T_{2}.
\end{equation}
The operators  $T_{\mu}$  are  defined by 
\begin{equation}
T_{\mu}\equiv \gamma_{\mu}\gamma_{5}\exp {(i\pi x^{\mu}/a)}  
\end{equation}
and  satisfy the relation
\begin{equation}
T_{\mu}T_{\nu} + T_{\nu}T_{\mu}=2\delta_{\mu\nu}
\end{equation}
with  $T_{\mu}^{\dagger} = T_{\mu} = T^{-1}_{\mu}$ for 
anti-hermitian $\gamma_{\mu}$. 
We denote the 16 operators by ${\cal T}_{n}, \ \ n=0\sim 15$, in 
the following 
with ${\cal T}_{0}=1$.
By recalling that the operator $T_{\mu}$ adds the  momentum 
$\pi/a$ to the 
fermion momentum $k_{\mu}$, we cover the entire Brillouin zone  
\begin{equation}
- \frac{\pi}{2a} \leq k_{\mu} <  \frac{3\pi}{2a}
\end{equation}
by the operation (2.7) starting with the free fermion defined in
\begin{equation}
- \frac{\pi}{2a} \leq k_{\mu} <  \frac{\pi}{2a}.
\end{equation}
The operators in (2.8) commute with $\gamma_{5}$, whereas those 
in (2.9) anti-commute with $\gamma_{5}$ and thus change the 
sign of  chiral charge,  reproducing the 15 species doublers for 
(2.6) with correct chiral charge assignment; 
$\sum_{n=0}^{15}(-1)^{n}\gamma_{5} =0$. 

One may define  the near continuum configurations by
the momentum $k_{\mu}$ carried by the fermion
\begin{equation}
- \frac{\pi}{2a}\epsilon \leq k_{\mu} \leq \frac{\pi}{2a}\epsilon
\end{equation}
or
\begin{equation}
- \frac{\pi}{2}\epsilon \leq ak_{\mu} \leq \frac{\pi}{2}\epsilon
\end{equation}
for sufficiently small $a$ and $\epsilon$ combined with the 
operation ${\cal T}_{n}$ in (2.8) and (2.9). 
To identify each species doubler clearly in the near continuum 
configurations, we also keep $r/a$ and $m_{0}/a$ {\em finite} for 
$a\rightarrow$ small [8], and the gauge fields are assumed to 
be sufficiently smooth. For these configurations, we can 
approximate the operator $D_{W}$ by
\begin{equation}
D_{W}= i\Dslash + M_{n} + O(\epsilon^{2}) + O(a(gA_{\mu})^{2})
\end{equation}
for each species doubler, where the mass parameters $M_{n}$  
stand for $M_{0}= - \frac{m_{0}}{a}$ and one of 
\begin{eqnarray}
&&\frac{2r}{a}-\frac{m_{0}}{a},\ \ (4,-1);\ \ \ 
\frac{4r}{a}-\frac{m_{0}}{a},\ \ (6,1)\nonumber\\
&&\frac{6r}{a}-\frac{m_{0}}{a},\ \ (4,-1);\ \ \ 
\frac{8r}{a}-\frac{m_{0}}{a},\ \ (1,1)
\end{eqnarray}
for $n=1\sim 15$. Here we denoted (multiplicity, chiral charge) 
in the bracket for species doublers. In (2.16) we used the 
relation valid for the configurations (2.15), for example, 
\begin{eqnarray}
D_{W}e^{ikx}&\equiv&\sum_{y}D_{W}(x,y)e^{iky}\nonumber\\
&=& [\sum_{\mu}\gamma^{\mu}\frac{\sin ak_{\mu}}{a} + 
\frac{r}{a}\sum_{\mu}(1 - \cos ak_{\mu}) - \frac{m_{0}}{a}]
e^{ikx}
\nonumber\\
&=&[\gamma^{\mu}k_{\mu}( 1 + O(\epsilon^{2})) 
+ \frac{r}{a} O(\epsilon^{2})-\frac{m_{0}}{a}]e^{ikx}
\end{eqnarray}
for vanishing gauge fields.

For the near continuum configurations, we thus have from (2.3)
\begin{equation}
D_{W}^{(2k+1)}=i(\Dslash)^{2k+1}+M_{n}^{(2k+1)}+
O(\epsilon^{2})
\end{equation}
where the mass parameters $M_{n}^{(2k+1)}$  
stand for 
\begin{equation}
M_{0}^{(2k+1)}\equiv-(\frac{m_{0}}{a})^{2k+1}
\end{equation}
and one of 
\begin{eqnarray}
&&(\frac{2r}{a})^{2k+1}-(\frac{m_{0}}{a})^{2k+1},\ \ (4,-1);\ \ \ 
(\frac{4r}{a})^{2k+1}-(\frac{m_{0}}{a})^{2k+1},\ \ (6,1)
\nonumber\\
&&(\frac{6r}{a})^{2k+1}-(\frac{m_{0}}{a})^{2k+1},\ \ (4,-1);\ \ \ 
(\frac{8r}{a})^{2k+1}-(\frac{m_{0}}{a})^{2k+1},\ \ (1,1)
\end{eqnarray}
for $n=1\sim 15$, in the same notation as in (2.17). 

To avoid the appearance of species doublers in $\gamma_{5}D$, we 
choose $M_{0}^{(2k+1)}<0$ and all other mass parameters 
$M_{n}^{(2k+1)}>0, \ \ n\neq0,$  namely
\begin{equation}
0<m_{0}<2r.
\end{equation}
The choice 
\begin{equation}
2m_{0}^{2k+1}=1
\end{equation}
normalizes properly the Dirac operator $H_{(2k+1)}$ in (2.2)
\begin{equation}
H_{(2k+1)}\simeq (i\gamma_{5}a\Dslash)^{2k+1}+\gamma_{5}
(i\gamma_{5}a\Dslash)^{2(2k+1)}
\end{equation}
in the near continuum configurations for all $|M_{n}|\rightarrow$
 large[1].

\section{Evaluation of the lattice Jacobian}
 
For an operator $O(x,y)$ defined on the lattice,
one may define 
\begin{equation}
O_{mn}\equiv \sum_{x,y}\phi_{m}^{\ast}(x)O(x,y)\phi_{n}(y),
\end{equation}
and the trace
\begin{eqnarray}
Tr O &=& \sum_{n}O_{nn}\nonumber\\
&=&\sum_{n}\sum_{x,y}\phi_{n}^{\ast}(x)O(x,y)\phi_{n}(y)
\nonumber\\
&=&\sum_{x}(\sum_{n,y}\phi_{n}^{\ast}(x)O(x,y)\phi_{n}(y)).
\end{eqnarray}
The local version of the trace (or anomaly) is then defined by 
\begin{equation}
tr O(x,x) \equiv\\ \sum_{n,y}\phi_{n}^{\ast}(x)O(x,y)\phi_{n}(y).
\end{equation}

For the operator of our interest, we have 
\begin{eqnarray}
&&tr\Gamma_{5}(x)\nonumber\\
&&=tr[\gamma_{5}-(\gamma_{5}aD)^{2k+1}]
\nonumber\\
&&=-tr(\gamma_{5}aD)^{2k+1}\nonumber\\
&&=-tr\frac{1}{2}\gamma_{5}[1+D_{W}^{(2k+1)}\frac{1}
{\sqrt{(D_{W}^{(2k+1)})^{\dagger}D_{W}^{(2k+1)}}}]\nonumber\\
&&=-tr\frac{1}{2}\gamma_{5}[D_{W}^{(2k+1)}\frac{1}
{\sqrt{(D_{W}^{(2k+1)})^{\dagger}D_{W}^{(2k+1)}}}]\\
&&= -\frac{1}{2}\sum_{n=0}^{15} 
tr\int^{\frac{\pi}{2a}}_{-\frac{\pi}{2a}}\frac{d^{4}k}
{(2\pi)^{4}}e^{-ikx}{\cal T}^{-1}_{n}\gamma_{5}D_{W}^{(2k+1)}
\frac{1}{\sqrt{(D_{W}^{(2k+1)})^{\dagger}D_{W}^{(2k+1)}}}
{\cal T}_{n}e^{ikx}\nonumber
\end{eqnarray}
where we used the plane wave basis defined in the domain (2.13) 
combined with the operation ${\cal T}_{n}$. 
In this calculation, we repeatedly used the relation
\begin{equation}
tr\gamma_{5}=0
\end{equation}
which is expected to be valid in lattice theory.
We also used a short hand notation 
\begin{equation}
Oe^{ikx}= \sum_{y}O(x,y)e^{iky}.
\end{equation}
There are various ways to evaluate the above trace (3.4).
We evaluate the trace (3.4) by following the procedure used 
for the overlap Dirac operator in Refs.[9][10]. Some of the 
basic papers of the lattice anomaly calculation are found in 
[11]-[14]. In this section we simplify the expression of the 
Jacobian, and its explicit evaluation is presented in the 
next section. 

\subsection{General analysis of the trace}

Our starting formula is (by using the momentum domain (2.12))
\begin{equation}
-\frac{1}{2}(\frac{1}{a})^{4}\int^{\frac{3\pi}{2}
}_{-\frac{\pi}{2}}\frac{d^{4}p}{(2\pi)^{4}}
tr\gamma_{5}\tilde{D}_{W}^{(2k+1)}(p)
\frac{1}{\sqrt{(\tilde{D}_{W}^{(2k+1)}(p))^{\dagger}
\tilde{D}_{W}^{(2k+1)}(p)}}
\end{equation}
with
\begin{eqnarray}
\tilde{D}_{W}^{(2k+1)}(p)&\equiv& (a^{2k+1})D_{W}^{(2k+1)}(p)
\nonumber\\
&=&i[i\sum_{\mu}\gamma^{\mu}\sin p_{\mu}+a\tilde{\Cslash}]^{2k+1}
+[r\sum_{\mu}(1-\cos p_{\mu})+a\tilde{B}]^{2k+1}\nonumber\\
&&-(m_{0})^{2k+1}
\end{eqnarray}
and we defined the integration variable
\begin{equation}
p_{\mu}= ak_{\mu}. 
\end{equation}
We used the definitions
\begin{equation}
e^{-ikx}a\Cslash e^{ikx}h(x)\equiv
[i\sum_{\mu}\gamma^{\mu}\sin ak_{\mu}+a\tilde{\Cslash}]h(x)
\end{equation}
and 
\begin{equation}
e^{-ikx}aBe^{ikx}h(x)\equiv[r\sum_{\mu}(1-\cos ak_{\mu})+
a\tilde{B}]h(x).
\end{equation}
for a sufficiently smooth function $h(x)$. In the following we
often omit writing $h(x)$.
Consequently,
\begin{eqnarray}
D_{W}^{(2k+1)}(k_{\mu})&\equiv& e^{-ikx}D_{W}^{(2k+1)}e^{ikx}
\nonumber\\
&=&i[i\sum_{\mu}\gamma^{\mu}\frac{\sin ak_{\mu}}{a}
+\tilde{\Cslash}]^{2k+1}
+[\frac{r}{a}\sum_{\mu}(1-\cos ak_{\mu})+ \tilde{B}]^{2k+1}
\nonumber\\
&&-(\frac{m_{0}}{a})^{2k+1}
\end{eqnarray}
and $D_{W}^{(2k+1)}(p)$ is defined by setting $k_{\mu}= p_{\mu}/a$
 in $D_{W}^{(2k+1)}(k_{\mu})$. 

In the continuum limit $a\rightarrow 0$ with $p_{\mu}=ak_{\mu}$
kept fixed,
the operator $\tilde{\Cslash}$ approaches 
\begin{equation}
\tilde{\Cslash}=\sum_{\mu}\gamma^{\mu}(\cos ak_{\mu}\partial_{\mu}
+ig\cos ak_{\mu}A_{\mu})+O(a)=\sum_{\mu}\gamma^{\mu}\cos p_{\mu}
D_{\mu}+O(a)
\end{equation}
and the leading term of $\tilde{B}$ is known to be[11]
\begin{equation}
\tilde{B}=-ir\sum_{\mu}\sin ak_{\mu} D_{\mu}+O(a)
=-ir\sum_{\mu}\sin p_{\mu}D_{\mu}+O(a)
\end{equation}
with the covariant derivative defined by
\begin{equation} 
D_{\mu}\equiv\partial_{\mu}+igA_{\mu}.
\end{equation}
 Note that the ``conventional naive continuum limit''
is defined by $a\rightarrow 0$ with $k_{\mu}$ kept fixed, instead 
of $p_{\mu}=ak_{\mu}$ being kept fixed as in the above limit. 

In the denominator of (3.7), one has a factor 
\begin{eqnarray}
&&(\tilde{D}_{W}^{(2k+1)}(p))^{\dagger}
\tilde{D}_{W}^{(2k+1)}(p)\nonumber\\
&&=[(i\sum_{\mu}\gamma^{\mu}\sin p_{\mu}+a\tilde{\Cslash})^{2}
]^{2k+1}\nonumber\\
&&+\{[r\sum_{\mu}(1-\cos p_{\mu})+a\tilde{B}]^{2k+1}
-(m_{0})^{2k+1}\}^{2}\nonumber\\
&&-i
[[i\sum_{\mu}\gamma^{\mu}\sin p_{\mu}+a\tilde{\Cslash}]^{2k+1},
 [r\sum_{\mu}(1-\cos p_{\mu})+a\tilde{B}]^{2k+1}]
\nonumber\\
&&=\{\sum_{\mu}(\sin p_{\mu}-ai\tilde{C}_{\mu})^{2}
+\frac{a^{2}}{4}[\gamma^{\mu},\gamma^{\nu}]
[\tilde{C}_{\mu},\tilde{C}_{\nu}]\}^{2k+1}
\nonumber\\
&&+\{[r\sum_{\mu}(1-\cos p_{\mu})+a\tilde{B}]^{2k+1}
-(m_{0})^{2k+1}\}^{2}\nonumber\\
&&-i
[[i\sum_{\mu}\gamma^{\mu}\sin p_{\mu}+a\tilde{\Cslash}]^{2k+1},
 [r\sum_{\mu}(1-\cos p_{\mu})+a\tilde{B}]^{2k+1}].
\end{eqnarray}
Note that the first two terms in this last expression commute 
with $\gamma_{5}$, while the last term anti-commutes with 
$\gamma_{5}$.
The last term of (3.16) is the interference term: From the 
structure of the commutator, one can confirm that 
it consists of terms with a factor
\begin{equation}
-ia^{2}\gamma^{\mu}[\tilde{C}_{\mu}, \tilde{B}]=
-ia^{2}gr\sum_{\mu,\nu}\gamma^{\mu}\cos p_{\mu}\sin p_{\nu}
F_{\mu\nu}+O(a^{3})
\end{equation}
and the $2k$ factors of 
$[i\sum_{\mu}\gamma^{\mu}\sin p_{\mu}+a\tilde{\Cslash}]$ 
and the $2k$ factors of 
$[r\sum_{\mu}(1-\cos p_{\mu})+a\tilde{B}]$.
We also note that 
\begin{equation}
[\tilde{C}_{\mu},\tilde{C}_{\nu}]=ig\cos p_{\mu}\cos p_{\nu}
F_{\mu\nu}+O(a).
\end{equation}
To simplify various expressions in the following, we define the 
variables 
\begin{equation}
c_{\mu}=\cos ak_{\mu}=\cos p_{\mu}, \ \ \ \ \ 
s_{\mu}=\sin ak_{\mu}=\sin p_{\mu}
\end{equation}
and 
\begin{equation}
\sslash=\sum_{\mu}\gamma^{\mu}\sin p_{\mu},\ \ \ \ \
s^{2}=\sum_{\mu}(s_{\mu})^{2}.
\end{equation}

\subsection{ Contribution of mass terms}

We now examine the integrand of (3.7) with only the ``mass terms''
in the numerator retained
\begin{equation}
\frac{1}{a^{4}}tr\gamma_{5}\{[r\sum_{\mu}(1-c_{\mu})
+a\tilde{B}]^{2k+1}
-(m_{0})^{2k+1}\}
\frac{1}{\sqrt{(\tilde{D}_{W}^{(2k+1)}(p))^{\dagger}
\tilde{D}_{W}^{(2k+1)}(p)}}.
\end{equation}
The numarator contains no $\gamma^{\mu}$'s. We expand the 
denominator in powers of the interference term, which contains
an odd number of $\gamma^{\mu}$'s. By remembering that the rest 
of the denominator factor contains an even number of 
$\gamma^{\mu}$'s and thus commute with $\gamma_{5}$, the odd
powers in the interference term in this expansion anti-commute
with $\gamma_{5}$ and thus vanish after taking the trace.
Only the even powers in the interference term could survive
the trace operation with $\gamma_{5}$. Since the interference 
term is of order $O(a^{2})$, only the zeroth order term and the 
second order term in the interference term are important in the 
limit $a\rightarrow 0$.

We can see that the second order term in the interference 
vanishes. Since the second order term is already of order 
$O(a^{4})$ because of (3.17),
one can set $a=0$ in all the remaining 
 $2k$ powers of 
$[i\sslash+a\tilde{\Cslash}]$ 
and the $2k$ powers of 
$[r\sum_{\mu}(1-c_{\mu})+a\tilde{B}]$. Namely these 
terms are replaced by 
$i\sslash$ 
and 
$r\sum_{\mu}(1-c_{\mu})$, respectively. Since these factors 
commute with each other, the interference term consists of 
a sum of terms of the structure
\begin{equation}
(i\sslash)^{l}
ia^{2}gr\sum_{\mu,\nu}\gamma^{\mu}c_{\mu}s_{\nu}
(i\sslash)^{m}
[r\sum_{\mu}(1-c_{\mu})]^{2k}
\end{equation}
with $l+m=2k$, if one uses (3.17): Thess terms are linear in 
$\gamma^{\mu}$ if one uses the relations
\begin{eqnarray}
&&(i\sslash)^{2}=
s^{2},\nonumber\\
&&(i\sslash)\gamma^{\nu}
(i\sslash)=
-2(\sslash)s_{\nu}
-s^{2}\gamma^{\nu}.
\end{eqnarray}
We thus have only two $\gamma^{\mu}$'s with order $O(a^{4})$
in the numerator of (3.21), which vanishes after the trace with 
$\gamma_{5}$. Note that the terms which contain $\gamma^{\mu}$'s
in the denominator is of order $O(a^{2})$ as in (3.16), and thus 
these terms cannot be used to supply extra $\gamma^{\mu}$'s.

We have thus established that only the zeroth order term 
in the interference survives in (3.21), namely, one can set 
the interference term to $0$ in the denominator. In this case, 
from the expression in (3.16) we see that the $\gamma^{\mu}$ 
factors appear only in the combination
\begin{equation}
\frac{a^{2}}{4}[\gamma^{\mu},\gamma^{\nu}]
[\tilde{C}_{\mu},\tilde{C}_{\nu}].
\end{equation}
The second power in this factor is just sufficient to survive
the trace operation and cancel $1/a^{4}$ in front of the 
integral. This means that we can set $a=0$ everywhere except 
in the prefactor in (3.24). The expression (3.21) is then 
replaced by 
\begin{equation}
\frac{1}{a^{4}}tr\gamma_{5}\{[r\sum_{\mu}(1-c_{\mu})]^{2k+1}
-(m_{0})^{2k+1}\}
\frac{1}{\sqrt{F_{(k)}}}
\end{equation}
where
\begin{eqnarray}
F_{(k)}&\equiv& \{ s^{2}
+ig\frac{a^{2}}{4}[\gamma^{\mu},\gamma^{\nu}]c_{\mu}
c_{\nu}F_{\mu\nu}
\}^{2k+1}
\nonumber\\
&&+\{[r\sum_{\mu}(1-c_{\mu})]^{2k+1}-(m_{0})^{2k+1}\}^{2}. 
\end{eqnarray}

\subsection{ Contribution of kinetic term}

Similarly, we analyze the integrand of (3.7) with the 
``kinetic term'' in the numerator
\begin{equation}
\frac{1}{a^{4}}tr\gamma_{5}\{i[i\sslash+a\tilde{\Cslash}]^{2k+1}\}
\frac{1}{\sqrt{(\tilde{D}_{W}^{(2k+1)}(p))^{\dagger}
\tilde{D}_{W}^{(2k+1)}(p)}}.
\end{equation}
Since the numerator is now odd in powers of $\gamma^{\mu}$, only 
the odd powers of the interference term could survive. The third 
power of the interference is $O(a^{6})$, and only the first 
power in the interference need to be analyzed. 

We first rewrite the numerator factor as 
\begin{eqnarray}
&&[i\sslash+a\tilde{\Cslash}]^{2k+1}\\
&&=\{\sum_{\mu}(s_{\mu}-ai\tilde{C}_{\mu})^{2}
+\frac{a^{2}}{4}[\gamma^{\mu},\gamma^{\nu}]
[\tilde{C}_{\mu},\tilde{C}_{\nu}]\}^{k}
[i\sslash+a\tilde{\Cslash}]\nonumber
\end{eqnarray}
which shows that we have only one $\gamma^{\mu}$ which is 
not multiplied by $a$. As we have already explained, the 
interference term in the denominator 
\begin{equation}
-i
[[i\sslash+a\tilde{\Cslash}]^{2k+1},
 [r\sum_{\mu}(1-c_{\mu})+a\tilde{B}]^{2k+1}]
\end{equation}
is written as a sum of terms with a single commutator
\begin{equation}
-ia^{2}\gamma^{\mu}[\tilde{C}_{\mu}, \tilde{B}]=
-ia^{2}gr\sum_{\mu,\nu}\gamma^{\mu}c_{\mu}s_{\nu}
F_{\mu\nu}+O(a^{3})
\end{equation}
multiplied by the $2k$ factors of 
$[i\sslash+a\tilde{\Cslash}]$ 
and the $2k$ factors of 
$[r\sum_{\mu}(1-c_{\mu})+a\tilde{B}]$. 
In such a term, if one exchanges the order of  
$[i\sslash+a\tilde{\Cslash}]$ 
and 
$[r\sum_{\mu}(1-c_{\mu})+a\tilde{B}]$, one generates 
another commutator as in (3.30). We then have a factor $a^{4}$
and thus we can set $a=0$ in all other terms in the integrand.
We recognize that such a term contains $\gamma^{\mu}$ only in 
the combination with $\sslash$
and the two factors of the above commutator (3.30). From this 
combination together with $\sslash$ in the numerator (3.28),
we cannot form a non-vanishing contraction with
the antisymmetric $\epsilon^{\mu\nu\alpha\beta}$ tensor. 

This means that we can write the interference term as a sum of 
terms of the structure
\begin{eqnarray}
&&[i\sslash+a\tilde{\Cslash}]^{l}
(-i)a^{2}gr\sum_{\mu,\nu}\gamma^{\mu}c_{\mu}s_{\nu}
F_{\mu\nu} 
[i\sslash+a\tilde{\Cslash}]^{m}
\nonumber\\
&&\times[r\sum_{\mu}(1-c_{\mu})+a\tilde{B}]^{2k}
\end{eqnarray}
where $l+m=2k$. If both of $l$ and $m$ are even, we can use 
\begin{equation}
[i\sslash+a\tilde{\Cslash}]^{2}
=\sum_{\mu}(s_{\mu}-ai\tilde{C}_{\mu})^{2}
+\frac{a^{2}}{4}[\gamma^{\mu},\gamma^{\nu}]
[\tilde{C}_{\mu},\tilde{C}_{\nu}]
\end{equation}
and  $\gamma^{\mu}$ always appears in the combination
$a\gamma^{\mu}$ except for the numerator term (3.28) which 
contains $\sslash$, and the 
commutator term (3.30), which contains $a^{2}\gamma_{\mu}$.
Such a combination could give rise to a non-vanishing result.
On the other hand, if both of  $l$ and $m$ are odd, one has 
to deal with a left-over term
\begin{eqnarray}
&&[i\sslash+a\tilde{\Cslash}]
(-i)a^{2}gr\sum_{\mu,\nu}\gamma^{\mu}c_{\mu}s_{\nu}
F_{\mu\nu} 
[i\sslash+a\tilde{\Cslash}]
\nonumber\\
&&=[i\sslash+a\tilde{\Cslash}]
(-2)(-i)a^{2}gr\sum_{\mu,\nu}c_{\mu}s_{\nu}
F_{\mu\nu}[is_{\mu}+a\tilde{C}_{\mu}]
\nonumber\\
&&+[i\sslash+a\tilde{\Cslash}]
(-i)a^{3}gr\sum_{\mu,\nu,\alpha}\gamma^{\alpha}\gamma^{\mu}
c_{\alpha}c_{\mu}s_{\nu}D_{\alpha}F_{\mu\nu}\nonumber\\
&& -\{\sum_{\mu}(s_{\mu}-ai\tilde{C}_{\mu})^{2}
+\frac{a^{2}}{4}[\gamma^{\mu},\gamma^{\nu}]
[\tilde{C}_{\mu},\tilde{C}_{\nu}]\}
(-i)a^{2}gr\sum_{\mu,\nu}\gamma^{\mu}c_{\mu}s_{\nu}
F_{\mu\nu}.\nonumber\\
&&
\end{eqnarray}
where we used $\gamma^{\mu}\gamma^{\alpha}
+\gamma^{\alpha}\gamma^{\mu}=-2\eta^{\mu\alpha}$.
The first term of this equation contains the factor 
$a^{2}[i\sslash+a\tilde{\Cslash}]$,
which should be replaced by 
$a^{2}i\sslash$ and should be 
combined with the factor $i\sslash$
in the numerator factor (together with 
$a^{2}[\gamma^{\mu},\gamma^{\nu}]
[\tilde{C}_{\mu},\tilde{C}_{\nu}]$ from other factors) to obtain 
a possible non-zero result. But such a term cannot make a 
non-vanishing contraction with $\epsilon^{\mu\nu\alpha\beta}$. 
The second term is of orfer $O(a^{3})$ and contains 3 
$\gamma^{\mu}$'s. If this term is combined with 
$[i\sslash+a\tilde{\Cslash}]$ in the numerator, it becomes of 
order $O(a^{5})$  due to (3.32). If it is  combined with 
the drivative operator such as $a\tilde{C}_{\mu}$ in (3.16), when 
commuting with other denominator factors, it becomes 
$O(a^{4})$; such a term contains $\sslash^{2}=-s^{2}$ if combined 
with $\sslash$ in the numerator and vanishes after trace with
$\gamma_{5}$. 
Thus we can set the 
first and second terms to $0$ in the above equation (3.33). 

Only the last term in (3.33) can survive: Among those surviving 
terms, the even $l$ terms and odd $l$ terms cancel pairwise 
except one term in the interference term. 

By this way, we can write the total interference term as
\begin{eqnarray}
&&-(2k+1)ia^{2}gr\sum_{\mu,\nu}\gamma^{\mu}c_{\mu}s_{\nu}
F_{\mu\nu}
\{s^{2}
+ig\frac{a^{2}}{4}[\gamma^{\mu},\gamma^{\nu}]c_{\mu}
c_{\nu}F_{\mu\nu}]\}^{k}\nonumber\\
&&\times[r\sum_{\mu}(1-c_{\mu})]^{2k}
\end{eqnarray}
where the factor $2k+1$ comes from the $2k+1$ powers 
of $[r\sum_{\mu}(1-c_{\mu})+a\tilde{B}]$. We also set 
$a=0$ in all the terms without $\gamma^{\mu}$ since this 
does not influence the surviving terms. We can also set $a=0$
in the numerator factor except for the combination
$a^{2}[\gamma^{\mu},\gamma^{\nu}]$.
Note that the order of $\gamma^{\mu}$ and 
$[\gamma^{\mu},\gamma^{\nu}]$ can be changed freely in the 
expansion of the denominator in powers of $a^{2}$, since 
the surviving terms are contracted with $\gamma_{5}$ to 
give rise to  $\epsilon^{\mu\nu\alpha\beta}$.

To summarize this tedious analysis, we can write the 
integrand with the ``kinetic'' term (3.27) as 
\begin{eqnarray}
&&\frac{1}{a^{4}}tr\gamma_{5}\{(2k+1)r(ig\frac{a^{2}}{4})
[\gamma^{\mu},\gamma^{\nu}]c_{\mu}c_{\nu}
F_{\mu\nu}(\sum_{\alpha}\frac{s_{\alpha}^{2}}{4c_{\alpha}})
\nonumber\\
&&\times\{s^{2}+ig\frac{a^{2}}{4}[\gamma^{\mu},\gamma^{\nu}]
c_{\mu}c_{\nu}F_{\mu\nu}] \}^{2k}
[r\sum_{\mu}(1-c_{\mu})]^{2k}\}\frac{1}{F_{(k)}^{3/2}}
\end{eqnarray}
where $F_{(k)}$ is defined in (3.26). In writing this final
expression we used the following sequence of rewriting
\begin{eqnarray}
\sslash\sum_{\mu,\nu}\gamma^{\mu}c_{\mu}s_{\nu}
F_{\mu\nu}&=&\sum_{\alpha}\sum_{\mu,\nu}\gamma^{\alpha}
\gamma^{\mu}c_{\mu}s_{\nu}s_{\alpha}
F_{\mu\nu}\nonumber\\
&=&\frac{1}{2}\sum_{\alpha}\sum_{\mu,\nu}[\gamma^{\alpha},
\gamma^{\mu}]c_{\mu}s_{\nu}s_{\nu}\delta_{\nu,\alpha}
F_{\mu\nu}\nonumber\\
&=&\frac{1}{2}\sum_{\mu,\nu}[\gamma^{\nu},
\gamma^{\mu}]c_{\mu}c_{\nu}(s_{\nu}s_{\nu}/c_{\nu})
F_{\mu\nu}\nonumber\\
&=&-\frac{1}{2}\sum_{\mu,\nu}[\gamma^{\mu},
\gamma^{\nu}]c_{\mu}c_{\nu}F_{\mu\nu}
(\sum_{\alpha}s_{\alpha}^{2}/
4c_{\alpha})
\end{eqnarray}
Namely, only the term with two $\gamma^{\mu}$'s contributes 
to the final result, and the odd term in $s_{\nu}$ after 
integration over the momentum vanishes. In the last step, we 
used the lattice hypercubic symmetry by taking into account the 
contraction with the $\epsilon^{\mu\nu\alpha\beta}$ symbol later.

\section{Formula for the chiral anomaly and parameter 
independence}

The basic formula for the chiral anomaly is given by 
(3.7), (3.25) and (3.35). The next step is to expand the 
integrand in the powers of\\ $(ig\frac{a^{2}}{4})
[\gamma^{\mu},\gamma^{\nu}]c_{\mu}c_{\nu}
F_{\mu\nu}$ and retain only the terms which contain
the second power of this factor. We then combine the 
expansion with the formula
\begin{equation}
tr\gamma_{5}\{(ig\frac{1}{4})
[\gamma^{\mu},\gamma^{\nu}]c_{\mu}c_{\nu}
F_{\mu\nu}\}^{2}
=(\prod_{\alpha=1}^{4}c_{\alpha})g^{2}
tr\epsilon^{\mu\nu\alpha\beta}F_{\mu\nu}F_{\alpha\beta}
\end{equation}
where $\epsilon^{1234}=1$.
We thus write only the coefficients of the factor
\begin{equation}
g^{2}tr\epsilon^{\mu\nu\alpha\beta}F_{\mu\nu}F_{\alpha\beta}
\end{equation}
in the following.

The contribution from the ``mass terms'' (3.25)
is given by
\begin{equation}
-\frac{1}{16}\int^{\frac{3\pi}{2}
}_{-\frac{\pi}{2}}\frac{d^{4}p}{(2\pi)^{4}}
(\prod_{\alpha=1}^{4}c_{\alpha})
\frac{(2k+1)M_{(2k+1)}[3(2k+1)(s^{2})^{4k}-4k(s^{2})^{2k-1}H]}
{H^{5/2}}
\end{equation}
where
\begin{eqnarray}
&&H\equiv(s^{2})^{2k+1}+M_{(2k+1)}^{2}\nonumber\\
&&M_{(2k+1)}\equiv[r\sum_{\mu}(1-c_{\mu})]^{2k+1}-m_{0}^{2k+1}.
\end{eqnarray}
The contribution from the ``kinetic term'' (3.35) is written as
\begin{eqnarray}
&&-\frac{1}{16}\int^{\frac{3\pi}{2}
}_{-\frac{\pi}{2}}\frac{d^{4}p}{(2\pi)^{4}}
(\prod_{\alpha=1}^{4}c_{\alpha})(2k+1)
\{r(\sum_{\beta}\frac{s_{\beta}^{2}}{c_{\beta}})
[r\sum_{\mu}(1-c_{\mu})]^{2k}\}\nonumber\\
&&\times
\{4k(s^{2})^{2k-1}H-3(2k+1)(s^{2})^{4k}\}
\frac{1}{H^{5/2}}.
\end{eqnarray}
Thus the total contribution is given by
\begin{eqnarray}
I_{2k+1}&=&-\frac{2k+1}{16}\int^{\frac{3\pi}{2}
}_{-\frac{\pi}{2}}\frac{d^{4}p}{(2\pi)^{4}}
(\prod_{\alpha=1}^{4}c_{\alpha})
\{M_{(2k+1)}-r(\sum_{\beta}\frac{s_{\beta}^{2}}{c_{\beta}})
[r\sum_{\mu}(1-c_{\mu})]^{2k}\}\nonumber\\
&&\times
\{3(2k+1)(s^{2})^{4k}-4k(s^{2})^{2k-1}H\}
\frac{1}{H^{5/2}}.
\end{eqnarray}

\subsection{Parameter independence}

To analyze the parameter independence of the coefficient
of the chiral anomaly, we follow the procedure in Refs.[9][10].
We first rewrite the integral in the domain 
$-\frac{\pi}{2}\leq p_{\mu}\leq \frac{3\pi}{2}$ to the integral
in the domain $-\frac{\pi}{2}\leq p_{\mu}\leq \frac{\pi}{2}$
by using the variables $s_{\mu}=\sin p_{\mu}$ as
\begin{equation}
\int^{\frac{3\pi}{2}
}_{-\frac{\pi}{2}}\frac{d^{4}p}{(2\pi)^{4}}
(\prod_{\alpha=1}^{4}c_{\alpha})
=\sum_{\epsilon_{\mu}=\pm}(\prod_{\mu}\epsilon_{\mu})
\int^{1}_{-1}\frac{d^{4}s}{(2\pi)^{4}}
\end{equation}
where the symbol $\epsilon_{\mu}$ takes care of the 16 
(would-be) species doublers
\begin{equation}
\epsilon_{\mu}=(\pm,\pm,\pm,\pm).
\end{equation}
The following formula is also valid
\begin{equation}
c_{\mu}=\epsilon_{\mu}(1-s_{\mu}^{2})^{1/2}.
\end{equation}
In this new notation, we have 
\begin{eqnarray}
&&H(s)\equiv(s^{2})^{2k+1}+M_{(2k+1)}^{2}\nonumber\\
&&M_{(2k+1)}\equiv[r\sum_{\mu}
(1-\epsilon_{\mu}(1-s_{\mu}^{2})^{1/2})]^{2k+1}-m_{0}^{2k+1}
\end{eqnarray}
and we evaluate
\begin{eqnarray}
&&(2k+1)H(s)+\frac{1}{5}H(s)^{7/2}\sum_{\mu}s_{\mu}\frac{\partial}
{\partial s_{\mu}}H(s)^{-5/2}\nonumber\\
&&=(2k+1)H(s)+\frac{1}{3}H(s)^{5/2}\sum_{\mu}s_{\mu}
\frac{\partial}{\partial s_{\mu}}H(s)^{-3/2}\nonumber\\
&&=-(2k+1)M_{(2k+1)}\{
-[r\sum_{\mu}
(1-\epsilon_{\mu}(1-s_{\mu}^{2})^{1/2})]^{2k+1}
+m_{0}^{2k+1}\nonumber\\
&&+[r\sum_{\mu}
(1-\epsilon_{\mu}(1-s_{\mu}^{2})^{1/2})]^{2k}r\sum_{\mu}
\epsilon_{\mu}s_{\mu}^{2}(1-s_{\mu}^{2})^{-1/2}\}
\nonumber\\
&&=(2k+1)M_{(2k+1)}\\
&&\times\{M_{(2k+1)}-[r\sum_{\mu}
(1-\epsilon_{\mu}(1-s_{\mu}^{2})^{1/2})]^{2k}r\sum_{\mu}
s_{\mu}^{2}\epsilon_{\mu}(1-s_{\mu}^{2})^{-1/2}\}\nonumber
\end{eqnarray}
which is a generalization of the identity discussed in [10].
By using these relations one can prove
\begin{eqnarray}
&&\frac{1}{2k+1}[\sum_{\mu}s_{\mu}\frac{\partial}
{\partial s_{\mu}}+4](\frac{(s^{2})^{4k}}{H(s)^{5/2}})
\nonumber\\
&&=-\frac{(s^{2})^{4k}}{H(s)^{5/2}}\nonumber\\
&&+\{M_{(2k+1)}-[r\sum_{\mu}
(1-\epsilon_{\mu}(1-s_{\mu}^{2})^{1/2})]^{2k}r\sum_{\mu}
s_{\mu}^{2}\epsilon_{\mu}(1-s_{\mu}^{2})^{-1/2}\}\nonumber\\
&&\times 5M_{(2k+1)}(\frac{(s^{2})^{4k}}{H(s)^{7/2}})
\end{eqnarray}
and
\begin{eqnarray}
&&\frac{1}{2k+1}[\sum_{\mu}s_{\mu}\frac{\partial}
{\partial s_{\mu}}+4](\frac{(s^{2})^{2k-1}}{H(s)^{3/2}})
\nonumber\\
&&=-\frac{(s^{2})^{2k-1}}{H(s)^{3/2}}\nonumber\\
&&+\{M_{(2k+1)}-[r\sum_{\mu}
(1-\epsilon_{\mu}(1-s_{\mu}^{2})^{1/2})]^{2k}r\sum_{\mu}
s_{\mu}^{2}\epsilon_{\mu}(1-s_{\mu}^{2})^{-1/2}\}\nonumber\\
&&\times 3M_{(2k+1)}(\frac{(s^{2})^{2k-1}}{H(s)^{5/2}}).
\end{eqnarray}
One can then show that 
\begin{eqnarray}
&&\frac{\partial I_{2k+1}}{\partial m_{0}^{2k+1}}
=\frac{2k+1}{16}\sum_{\epsilon_{\mu}=\pm}(\prod_{\mu}
\epsilon_{\mu})\int^{1}_{-1}\frac{d^{4}s}{(2\pi)^{4}}\nonumber\\
&&\times\{3(2k+1)[\frac{(s^{2})^{4k}}{H^{5/2}}
-5[M_{(2k+1)}-r(\sum_{\beta}\frac{s_{\beta}^{2}}
{c_{\beta}})
[r\sum_{\mu}(1-c_{\mu})]^{2k}]M_{(2k+1)}
\frac{(s^{2})^{4k}}{H^{7/2}}]\nonumber\\
&&-4k[\frac{(s^{2})^{2k-1}}{H^{3/2}}
-3[M_{(2k+1)}-r(\sum_{\beta}\frac{s_{\beta}^{2}}
{c_{\beta}})
[r\sum_{\mu}(1-c_{\mu})]^{2k}]M_{(2k+1)}
\frac{(s^{2})^{2k-1}}{H^{5/2}}]\}\nonumber\\
&&=\frac{2k+1}{16}\sum_{\epsilon_{\mu}=\pm}(\prod_{\mu}
\epsilon_{\mu})\int^{1}_{-1}\frac{d^{4}s}{(2\pi)^{4}}\nonumber\\
&&\times\{[-3(\sum_{\mu}s_{\mu}\frac{\partial}
{\partial s_{\mu}}+4)(\frac{(s^{2})^{4k}}{H(s)^{5/2}})]
\nonumber\\
&&+\frac{4k}{2k+1}[(\sum_{\mu}s_{\mu}\frac{\partial}
{\partial s_{\mu}}+4)(\frac{(s^{2})^{2k-1}}{H(s)^{3/2}})]\}
\end{eqnarray}
Similarly, we can show the relations by using (4.11)
\begin{eqnarray}
&&\frac{1}{2k+1}[\sum_{\mu}s_{\mu}\frac{\partial}
{\partial s_{\mu}}+4](\frac{[r\sum_{\mu}(1-c_{\mu})]^{2k+1}
(s^{2})^{4k}}{H(s)^{5/2}})
\nonumber\\
&&=-\frac{\{[r\sum_{\mu}(1-c_{\mu})]^{2k+1}-[r\sum_{\mu}
(1-c_{\mu})]^{2k}r\sum_{\mu}
\epsilon_{\mu}s_{\mu}^{2}(1-s_{\mu}^{2})^{-1/2}\}(s^{2})^{4k}}
{H(s)^{5/2}}\nonumber\\
&&+\{M_{(2k+1)}-[r\sum_{\mu}
(1-c_{\mu})]^{2k}r\sum_{\mu}
s_{\mu}^{2}\epsilon_{\mu}(1-s_{\mu}^{2})^{-1/2}\}\nonumber\\
&&\times 5M_{(2k+1)}(\frac{[r\sum_{\mu}(1-c_{\mu})]^{2k+1}
(s^{2})^{4k}}{H(s)^{7/2}})
\end{eqnarray}
and
\begin{eqnarray}
&&\frac{1}{2k+1}[\sum_{\mu}s_{\mu}\frac{\partial}
{\partial s_{\mu}}+4](\frac{[r\sum_{\mu}(1-c_{\mu})]^{2k+1}
(s^{2})^{2k-1}}{H(s)^{3/2}})
\nonumber\\
&&=-\frac{\{[r\sum_{\mu}(1-c_{\mu})]^{2k+1}-[r\sum_{\mu}
(1-c_{\mu})]^{2k}r\sum_{\mu}
\epsilon_{\mu}s_{\mu}^{2}(1-s_{\mu}^{2})^{-1/2}\}(s^{2})^{2k-1}}
{H(s)^{3/2}}\nonumber\\
&&+\{M_{(2k+1)}-[r\sum_{\mu}
(1-c_{\mu})]^{2k}r\sum_{\mu}
s_{\mu}^{2}\epsilon_{\mu}(1-s_{\mu}^{2})^{-1/2}\}\nonumber\\
&&\times 3M_{(2k+1)}(\frac{[r\sum_{\mu}(1-c_{\mu})]^{2k+1}
(s^{2})^{2k-1}}{H(s)^{5/2}})
\end{eqnarray}
where $c_{\mu}=\epsilon_{\mu}(1-s_{\mu}^{2})^{1/2}$ is understood.
Using these relations, we have
\begin{eqnarray}
&&r^{2k+1}\frac{\partial I_{2k+1}}{\partial r^{2k+1}}
=-\frac{2k+1}{16}\sum_{\epsilon_{\mu}=\pm}(\prod_{\mu}
\epsilon_{\mu})\int^{1}_{-1}\frac{d^{4}s}{(2\pi)^{4}}\nonumber\\
&&\times\{3(2k+1)\{\frac{\{[r\sum_{\mu}(1-c_{\mu})]^{2k+1}-
[r\sum_{\mu}
(1-c_{\mu})]^{2k}r\sum_{\mu}
\epsilon_{\mu}s_{\mu}^{2}(1-s_{\mu}^{2})^{-1/2}\}(s^{2})^{4k}}
{H(s)^{5/2}}\nonumber\\
&&-[M_{(2k+1)}-[r\sum_{\mu}
(1-c_{\mu})]^{2k}r\sum_{\mu}
s_{\mu}^{2}\epsilon_{\mu}(1-s_{\mu}^{2})^{-1/2}]\nonumber\\
&&\times 5M_{(2k+1)}(\frac{[r\sum_{\mu}(1-c_{\mu})]^{2k+1}
(s^{2})^{4k}}{H(s)^{7/2}})
\}\nonumber\\
&&-4k\{\frac{\{[r\sum_{\mu}(1-c_{\mu})]^{2k+1}-[r\sum_{\mu}
(1-c_{\mu})]^{2k}r\sum_{\mu}
\epsilon_{\mu}s_{\mu}^{2}(1-s_{\mu}^{2})^{-1/2}\}(s^{2})^{2k-1}}
{H(s)^{3/2}}\nonumber\\
&&-[M_{(2k+1)}-[r\sum_{\mu}
(1-c_{\mu})]^{2k}r\sum_{\mu}
s_{\mu}^{2}\epsilon_{\mu}(1-s_{\mu}^{2})^{-1/2}]\nonumber\\
&&\times 3M_{(2k+1)}(\frac{[r\sum_{\mu}(1-c_{\mu})]^{2k+1}
(s^{2})^{2k-1}}{H(s)^{5/2}})\}\}\nonumber\\
&&=\frac{2k+1}{16}\sum_{\epsilon_{\mu}=\pm}(\prod_{\mu}
\epsilon_{\mu})\int^{1}_{-1}\frac{d^{4}s}{(2\pi)^{4}}\nonumber\\
&&\times\{[-3(\sum_{\mu}s_{\mu}\frac{\partial}
{\partial s_{\mu}}+4)(\frac{[r\sum_{\mu}(1-c_{\mu})]^{2k+1}
(s^{2})^{4k}}{H(s)^{5/2}})]
\nonumber\\
&&+\frac{4k}{2k+1}[(\sum_{\mu}s_{\mu}\frac{\partial}
{\partial s_{\mu}}+4)(\frac{[r\sum_{\mu}(1-c_{\mu})]^{2k+1}
(s^{2})^{2k-1}}{H(s)^{3/2}})]\}.
\end{eqnarray}
The integrand becomes singular only if the following two relations
simultaneously hold
\begin{equation}
s^{2}=0,\ \ \ \ \ 
[r\sum_{\mu}
(1-\epsilon_{\mu}(1-s_{\mu}^{2})^{1/2})]^{2k+1}-m_{0}^{2k+1}=0
\end{equation}
namely, only when $m_{0}/r=0,\ \ 2,\ \ 4,\ \ 6,\ \ 8$. We are working
in the physical region
\begin{equation}
0< m_{0}< 2r
\end{equation}
and thus the above integrals are regular, and we have from
(4.14) and (4.17)
\begin{equation}
\frac{\partial I_{2k+1}}{\partial m_{0}^{2k+1}}=
\frac{\partial I_{2k+1}}{\partial r^{2k+1}}=0
\end{equation}
after partial integration. It can be confirmed that boundary 
terms at $s_{\mu}=\pm 1$ give vanishing contributions after 
a summation over $\sum_{\epsilon_{\mu}=\pm}$. This shows that 
the coefficient of 
the anomaly is stable under a smooth variation of the parameters
 $r$ and $m_{0}$, which is expected for a topological quantity 
such as the chiral anomaly.

\subsection{Explicit evaluation of the chiral anomaly}

Since the coefficient of the anomaly is independent of the 
parameters $r$ and $m_{0}$, we evaluate the anomaly in the limit
where both of $r$ and $m_{0}$ go to $0$. To be precise
we introduce an auxiliary parameter $a$, and take a limit
$a\rightarrow 0$ with both of  
\begin{equation}
\frac{r}{a}, \ \ \ \ \ \frac{m_{0}}{a}
\end{equation}
kept {\em fixed} in the physical region (4.19). The parameter $a$
 plays the role of an 
effective lattice spacing, though our formulas (4.3) and (4.5) 
are derived in the limit of the lattice spacing $a=0$.
 
There are various ways to evaluate the coefficient of the anomaly
, and one of these methods is given in [10] in the analysis of 
the overlap operator with $k=0$. We present a 
calculation which reveals a close connnection with the naive
continuum limit. We first observe that the contribution of 
the ``kinetic'' term (4.5), which arises from the interference
term in the denominator, vanishes in the above limit (4.21)
\footnote{
This property has been used in the treatment of the overlap 
operator in Refs.[7] and [15].}. 
To show this we examine
\begin{eqnarray}
&&-\frac{1}{16}\sum_{\epsilon_{\mu}=\pm}
(\prod_{\mu}\epsilon_{\mu})
\int^{1}_{-1}\frac{d^{4}s}{(2\pi)^{4}}
\{r(\sum_{\beta}\frac{s_{\beta}^{2}}{c_{\beta}})
[r\sum_{\mu}(1-c_{\mu})]^{2k}\}\nonumber\\
&&\times
\{4k(s^{2})^{2k-1}H-3(2k+1)(s^{2})^{4k}\}
\frac{1}{H^{5/2}}.
\end{eqnarray}
where $c_{\mu}=\epsilon_{\mu}(1-s_{\mu}^{2})^{1/2}$ is understood.
We define the integration domain
\begin{equation}
-\epsilon\leq s_{\mu}\leq \epsilon
\end{equation}
for all $\mu$ with a sufficiently small but finite $\epsilon$.
Since $s^{2}>0$ and the denominator of the integrand is regular 
for the integration domain {\em outside} the above domain, the 
integral outside the domain (4.23) vanishes in the limit 
$a\rightarrow 0$. 
Note that the denominator of 
$\sum_{\beta}\frac{s_{\beta}^{2}}{c_{\beta}}$ does not cause any
divergence in the integral (4.22).
In fact one can even take $\epsilon\rightarrow 0$ 
in such a manner that 
\begin{equation}
a/\epsilon^{l}\rightarrow 0
\end{equation}
for a suitable fixed positive integer $l$. This is because the 
integral {\em outside} the domain (4.23) vanishes at least 
linearly in $a$, and thus one can let $\epsilon\rightarrow 0$ 
simultaneously with the above constraint, where the denominator 
$\epsilon^{l}$ stands for the possible infrared singularity in 
this calculational procedure.

We thus examine the remaining integral
\begin{eqnarray}
&&-\frac{1}{16}\sum_{\epsilon_{\mu}=\pm}
(\prod_{\mu}\epsilon_{\mu})
\int^{\epsilon}_{-\epsilon}\frac{d^{4}s}{(2\pi)^{4}}
\{r(\sum_{\beta}\frac{s_{\beta}^{2}}{c_{\beta}})
[r\sum_{\mu}(1-c_{\mu})]^{2k}\}\nonumber\\
&&\times
\{4k(s^{2})^{2k-1}H-3(2k+1)(s^{2})^{4k}\}
\frac{1}{H^{5/2}}.
\end{eqnarray}
Since $|c_{\mu}|\simeq 1$ in the above integral, we can ignore 
the variation of $c_{\mu}$. If one rescales the integration 
variable $s_{\mu}=as_{\mu}^{\prime}$ and defines
\begin{equation}
r^{\prime}=r/a, \ \ \ \ \ m_{0}^{\prime}=m_{0}/a
\end{equation}
the above integral is written as
\begin{eqnarray}
&&-\frac{1}{16}\sum_{\epsilon_{\mu}=\pm}
(\prod_{\mu}\epsilon_{\mu})
\int^{\epsilon/a}_{-\epsilon/a}\frac{d^{4}s^{\prime}}{(2\pi)^{4}}
\{r^{\prime}(\sum_{\beta}\frac{s_{\beta}^{2}}{c_{\beta}})
[r^{\prime}\sum_{\mu}(1-c_{\mu})]^{2k}\}\nonumber\\
&&\times
\{4k((s^{\prime})^{2})^{2k-1}H-3(2k+1)((s^{\prime})^{2})^{4k}\}
\frac{1}{H^{5/2}}
\end{eqnarray}
where $H(s^{\prime})$ is parametrized by $r^{\prime}$ and 
$m_{0}^{\prime}$, which are kept fixed in the limit 
$a\rightarrow 0$. Note that the factor 
$(\sum_{\beta}\frac{s_{\beta}^{2}}{c_{\beta}})$ in the numerator,
 which is written
in terms of the original variables, is $O(\epsilon^{2})$.
The above integral is convergent in this limit $a\rightarrow0$
and of order $O(\epsilon^{2})$, and thus it can be made 
arbitrarily small. We can even make it vanish precisely by taking
 the limit (4.24).
We can thus ignore the contribution of the ``kinetic'' term, 
which arises from the  
interference term in the denominator, in the above limit (4.21)
.

We now come to the main contribution of the ``mass terms'' in 
(4.3).
It turns out to be more convenient to go back to (3.25),
which gives rise to (4.3). If one uses the notation of (3.4)
instead of (3.7), (3.25) is written as 
\begin{equation}
\sum_{n=0}^{15}(\frac{-1}{2})(-1)^{n}\frac{1}{a^{4}}
\int^{\pi/2}_{-\pi/2}\frac{d^{4}p}{(2\pi)^{4}}   
tr\gamma_{5}\{[r\sum_{\mu}(1\pm c_{\mu})]^{2k+1}
-(m_{0})^{2k+1}\}
\frac{1}{\sqrt{F_{(k)}(n,p_{\mu})}}
\end{equation}
where
\begin{eqnarray}
F_{(k)}(n,p_{\mu})&\equiv& \{(i\sslash+\sum_{\mu}a\gamma^{\mu}
c_{\mu}
D_{\mu})^{2}\}^{2k+1}
\nonumber\\
&&+\{[r\sum_{\mu}(1\pm c_{\mu})]^{2k+1}-(m_{0})^{2k+1}\}^{2}. 
\end{eqnarray}
The summation runs over the 16 would-be species doublers and the 
factor $r\sum_{\mu}(1\pm c_{\mu})$ arises from each momentum 
domain. For later convenience, we modified the denominator factor
 in (3.25) by replacing 
$s^{2}+ig\frac{a^{2}}{4}[\gamma^{\mu},\gamma^{\nu}]c_{\mu}
c_{\nu}F_{\mu\nu} $ with $(i\sslash+\sum_{\mu}a\gamma^{\mu}c_{\mu}
D_{\mu})^{2}$ , but this does not change the result as was 
explained in detail in the passage from (3.21) to (3.25) in 
Section 3.2.

In the present integral, we can also define the domain
\begin{equation}
-\epsilon\leq p_{\mu}\leq \epsilon
\end{equation}
for arbitrarily small but finite $\epsilon$. One can again
confirm that the integral {\em outside} this domain vanishes 
at least linearly in $a$ for $a\rightarrow 0$. This is because 
we retain the second order term in 
$a^{2}[\gamma^{\mu},\gamma^{\nu}]F_{\mu\nu}$ to survive the trace
with $\gamma_{5}$, and this cancels the factor $1/a^{4}$ in front
of the integral. The resulting integral is finite {\em outside} 
the above domain, and it vanishes at least linearly in $a$ in 
the limit (4.21). We can also let $\epsilon\rightarrow 0$
as in (4.24).

We thus examine (4.28) inside the domain (4.30)
\begin{eqnarray}
&&\sum_{n=0}^{15}(\frac{-1}{2})(-1)^{n}\frac{1}{a^{4}}
\int^{\epsilon}_{-\epsilon}\frac{d^{4}p}{(2\pi)^{4}}   
tr\gamma_{5}
\frac{[r\sum_{\mu}(1\pm c_{\mu})]^{2k+1}
-(m_{0})^{2k+1}}{\sqrt{F_{(k)}(n,p_{\mu})}}\nonumber\\
&&=\sum_{n=0}^{15}(\frac{-1}{2})(-1)^{n}
\int^{\epsilon/a}_{-\epsilon/a}\frac{d^{4}k}{(2\pi)^{4}}   
tr\gamma_{5}
\frac{[r\sum_{\mu}(1\pm \cos ak_{\mu})]^{2k+1}
-(m_{0})^{2k+1}}{\sqrt{F_{(k)}(n,ak_{\mu})}}
\end{eqnarray}
For sufficiently small $\epsilon$, we have 
\begin{eqnarray}
&&[r\sum_{\mu}(1\pm \cos ak_{\mu})]^{2k+1}/a^{2k+1}
-(m_{0})^{2k+1}/a^{2k+1}=M_{n}^{(2k+1)} + O(\epsilon^{2})
\nonumber\\
&&F_{(k)}(n,ak_{\mu})/a^{2(2k+1)}=
(i\kslash+\Dslash)^{2(2k+1)}+(M_{n}^{(2k+1)})^{2}+O(\epsilon^{2})
\end{eqnarray}
where the mass parameters $M_{n}^{(2k+1)}$ are defined in 
(2.20) and (2.21).
 The above integral in the limit $a\rightarrow0$ with (4.24) 
approaches 
\begin{eqnarray}
&&-\frac{1}{2} \sum_{n=0}^{15}(-1)^{n}tr\int^{\infty}_{-\infty}
 \frac{d^{4}k}{(2\pi)^{4}}\gamma_{5}
\frac{M_{n}^{(2k+1)}}
{\sqrt{([i\kslash +\Dslash]^{2})^{2k+1}
+(M_{n}^{(2k+1)})^{2}}}\nonumber\\
&=&\frac{1}{2}tr \int^{\infty}_{-\infty}\frac{d^{4}k}
{(2\pi)^{4}}\gamma_{5}
\frac{1}{\sqrt{((i\kslash+\Dslash)^{2}/\hat{M}_{0}^{2})^{2k+1}+1}}
\nonumber\\
&&-\frac{1}{2}\sum_{n=1}^{15}(-1)^{n}tr \int^{\infty}_{-\infty}
\frac{d^{4}k}{(2\pi)^{4}}\gamma_{5}\frac{1}
{\sqrt{((i\kslash+\Dslash)^{2}/\hat{M}_{n}^{2})^{2k+1}+1}}
\end{eqnarray}
by recalling $\hat{M}_{0}<0$, and 
$\hat{M}_{n}^{2}\equiv[(M_{n}^{(2k+1)})^{2}]^{1/(2k+1)}$.  
The sum of integrals in (4.33) gives rise to the 
anomaly for all $\hat{M}_{n}^{2}\rightarrow \infty$ in the 
final stage:[15] 
\begin{equation}
\lim_{M\rightarrow \infty}tr \int^{\infty}_{-\infty}
\frac{d^{4}k}{(2\pi)^{4}}e^{-ikx}\gamma_{5}
f(\frac{\Dslash^{2}}{M^{2}})e^{ikx}
=\frac{g^{2}}{32\pi^{2}}tr\epsilon^{\mu\nu\alpha\beta}
F_{\mu\nu}F_{\alpha\beta}. 
\end{equation}
Here we defined 
\begin{equation}
f(x)=\frac{1}{\sqrt{x^{2k+1}+1}}
\end{equation}
which satisfies 
\begin{eqnarray}
&& f(0)=1,\ \ f(\infty)=0,\nonumber\\
&& f^{\prime}(x)x|_{x=0} = f^{\prime}(x)x|_{x=\infty} = 0.
\end{eqnarray}
The left-hand side of (4.34) is known to be independent of the 
choice of $f(x)$ which satisfies the mild condition (4.36)[6].

By combining (1.15), (3.4) and (4.34), we recover the 
Atiyah-Singer index theorem ( in continuum $R^{4}$ space)[16][17]
\begin{equation}
n_{+}-n_{-}=\int dx \frac{g^{2}}{32\pi^{2}}
tr\epsilon^{\mu\nu\alpha\beta}
F_{\mu\nu}F_{\alpha\beta}
\end{equation}
in the (naive) continuum limit for any operator in (1.1), if 
one follows  the construction of $\gamma_{5}D$ in [1].  

\section{Discussion}

We have shown in an explicit and elementary manner that a new 
class of lattice Dirac operators proposed in [1] satisfy the 
correct anomaly relation. We have in particular shown that the 
anomaly coefficient is independent of a small variation in the 
parameters $r$ and $m_{0}$, which characterize these Dirac 
operators. This is in agreement with the more general but formal 
analysis in [1].

The Dirac operator in [1] is defined in a somewhat
indirect manner as
\begin{equation}
\det H=(\det H_{(2k+1)})^{1/(2k+1)}
\end{equation}
as in (A.9) in Appendix.
This definition is sufficient for the non-perturbative 
analysis, and as we have shown in this paper, it is also
sufficient to evaluate the chiral anomaly explicitly.
However, the perturbative treatment of these general class
of Dirac operators is not well understood yet.
It would therefore be interesting to extend the analyses
in [14][18], for example, to the present general class of 
operators.

As for the locality issue of the present class of operators,
the fact that the anomaly calculation in a naive continuum 
limit makes sense suggests that there is a certain range of 
gauge field configurations which make these operators local.  
The fact that we take a 2k+1th root of
$H_{(2k+1)}$ in (5.1) by itself may not spoil much of the 
locality, since the eigenfunctions in (A.17) are defined in terms 
of $H_{(2k+1)}$ and thus they may reflect the locality 
properties of $H_{(2k+1)}$, which may not differ qualitatively 
from those of the overlap operator[19][20]. In any case, a direct 
analysis of this locality issue is left as an important problem. 
   
Our construction of the Dirac operator (5.1)
on the basis of the defining algebraic relation (1.1) suggests
that we obtain better chiral properties if one 
increases the parameter $k$. An intuitive argument for this 
expectation is that the right-hand side 
$2a^{2k+1}(\gamma_{5}D)^{2k+2}$
of the algebra (1.1), which breaks chiral symmetry, 
becomes more irrelevant for larger $k$ in the sense of 
renormalization group. At the same time, however, our construction
 requires a larger lattice for larger $k$ since the basic 
operator appearing in our construction spreads over far apart 
lattice points for large $k$.
 To maintain the locality of the Dirac operator, we need to take 
a smaller lattice spacing for larger $k$.

As for the chiral fermions[21][22], the present calculation
of anomaly is readily extended to the evaluation of the fermion 
number anomaly of chiral theory and also to the so-called 
covariant form of non-Abelian anomalies[6] in the 
continuum limit. But the construction of chiral fermion
theory at finite lattice spacing is a challenging un-solved 
problem not only in our general Dirac operators but also
in the original overlap operator[3].  

T-W. Chiu has recently informed us that a numerical study of  
some of basic properties of the operator $\gamma_{5}D$ with 
$k=1$, such as index theorem, chiral anomaly and the propagator,
is in progress[24].
\\

One of us (KF) thanks Ting-Wai Chiu for numerous helpful 
discussions from the very beginning of the present 
investigation.  

\appendix

\section{Basic construction of general Dirac operators}

We start with (1.1) written in the form[1]
\begin{equation}
H\gamma_{5}+\gamma_{5}H=2H^{2k+2}
\end{equation}
or equivalently
\begin{equation}
\Gamma_{5}H+H\Gamma_{5}=0
\end{equation}
where $H=a\gamma_{5}D$ and $\Gamma_{5}=\gamma_{5}-H^{2k+1}$.
This algebraic relation implies that
\begin{equation}
\gamma_{5}H^{2}=[\gamma_{5}H+H\gamma_{5}]H-
H[\gamma_{5}H+H\gamma_{5}]+H^{2}\gamma_{5}=H^{2}\gamma_{5}.
\end{equation}
Namely, the algebraic relation (A.1) is equivalent to the two 
relations
\begin{eqnarray}
&&H^{2k+1}\gamma_{5}+\gamma_{5}H^{2k+1}=2H^{2(2k+1)},\nonumber\\
&&\gamma_{5}H^{2}-H^{2}\gamma_{5}=0.
\end{eqnarray}
If one defines $H_{(2k+1)}\equiv H^{2k+1}$, the first relation 
of (A.4) becomes 
\begin{equation}
H_{(2k+1)}\gamma_{5}+\gamma_{5}H_{(2k+1)}=2H_{(2k+1)}^{2}
\end{equation}
with $\Gamma_{5}=\gamma_{5}-H_{(2k+1)}$, 
which is identical to the conventional Ginsparg-Wilson relation
with $k=0$ in (1.1). We utilize this property to construct a 
solution to (A.1).

The physical condition for the operator $H$ in (A.1) in the 
near continuum configuration is 
\begin{equation}
H\simeq \gamma_{5}ai\Dslash+\gamma_{5}(\gamma_{5}ai\Dslash)^{2k+2}
\end{equation}
where the first term stands for the leading term in chiral
symmetric terms, and the second term stands for the leading
term in chiral symmetry breaking terms. 
Thus $H_{(2k+1)}$ should satisfy
\begin{eqnarray}
H_{(2k+1)}&\simeq&[\gamma_{5}ai\Dslash
+\gamma_{5}(\gamma_{5}ai\Dslash)^{2k+2}]^{2k+1}\nonumber\\
&\simeq&(\gamma_{5}ai\Dslash)^{2k+1}
+\gamma_{5}(\gamma_{5}ai\Dslash)^{2(2k+1)}
\end{eqnarray}
as can be confirmed by noting $\gamma_{5}\Dslash
+\Dslash\gamma_{5}=0$. Here only the leading terms in chiral 
symmetric and chiral symmetry breaking terms, respectively, are 
written.

One can thus construct a solution for $H_{(2k+1)}$ by
\begin{equation}
H_{(2k+1)}=\frac{1}{2}\gamma_{5}[1+\gamma_{5}H_{W}^{(2k+1)}
\frac{1}{\sqrt{H_{W}^{(2k+1)}H_{W}^{(2k+1)}}}]
\end{equation}
in terms of the hermitian $H_{W}^{(2k+1)}\equiv\gamma_{5}
D_{W}^{(2k+1)}=(H_{W}^{(2k+1)})^{\dagger}$. The operator 
$D_{W}^{(2k+1)}$ is defined in (2.3).
The physical condition (A.7) is satisfied by (2.19), as was noted 
in the body of the text.

We now discuss how to reconstruct $H$, which satisfies (A.1),
 from  $H_{(2k+1)}$ defined
above. The basic idea is to define in the representation 
where $H_{(2k+1)}$ is diagonal
\begin{equation}
H=(H_{(2k+1)})^{1/(2k+1)}
\end{equation}
in such a manner that $H$ thus obtained satisfies the second 
constraint in (A.4). For this purpose, we first recall the 
essence of the general representation of the algebra (A.1)[1]. 
\\

If one defines the eigenvalue problem
\begin{equation}
H_{(2k+1)}\phi_{n}=(a\lambda_{n})^{2k+1}\phi_{n},\ \ \ 
(\phi_{n},\phi_{n})=1
\end{equation}
one can classify the eigenstates into the 3 classes:\\
(i)\ $n_{\pm}$ (``zero modes''),\\
\begin{equation}
H_{(2k+1)}\phi_{n}=0, \ \ \gamma_{5}\phi_{n} = \pm \phi_{n},
\end{equation}
(ii)\ $N_{\pm}$ (``highest states''), \\
\begin{equation}
H_{(2k+1)}\phi_{n}= \pm\phi_{n}, \ \ \
\gamma_{5}\phi_{n} = \pm \phi_{n},\ \ \ respectively,
\end{equation}
(iii)``paired states'' with $0 < |(a\lambda_{n})^{2k+1}| < 1$,
\begin{equation}
H_{(2k+1)}\phi_{n}= (a\lambda_{n})^{2k+1}\phi_{n}, \ \ \ 
H_{(2k+1)}(\Gamma_{5}\phi_{n})
= - (a\lambda_{n})^{2k+1}(\Gamma_{5}\phi_{n}).
\end{equation}
where
\begin{equation}
\Gamma_{5}=\gamma_{5}-H_{(2k+1)}.
\end{equation}
Note that $\Gamma_{5}(\Gamma_{5}\phi_{n})\propto \phi_{n}$ for 
$0<|(a\lambda_{n})^{2k+1}|<1$.\\

We have a chirality sum rule[23]
\begin{equation}
n_{+}+N_{+}=n_{-}+N_{-}  
\end{equation} 
where $N_{\pm}$ stand for the number of ``highest states''
in the classification (ii) above.

All the states $\phi_{n}$ with 
$0<|(a\lambda_{n})^{2k+1}|<1$, 
which appear pairwise with $(a\lambda_{n})^{2k+1}= \pm 
|(a\lambda_{n})^{2k+1}|$, 
can be normalized to satisfy the relations
\begin{eqnarray}
\Gamma_{5}\phi_{n}&=&
[1-(a\lambda_{n})^{2(2k+1)}]^{1/2}\phi_{-n},
\nonumber\\
\gamma_{5}\phi_{n}&=&(a\lambda_{n})^{2k+1}\phi_{n}+
[1-(a\lambda_{n})^{2(2k+1)}]^{1/2}\phi_{-n},
\end{eqnarray}
where $\phi_{-n}$ stands for the eigenstate with an eigenvalue
opposite to that of $\phi_{n}$.\\

We can define the solution $H$ of (A.1) {\em operationally} by
\begin{equation}
H\phi_{n}\equiv a\lambda_{n}\phi_{n}
\end{equation}
by using the {\em same set} of eigenfunctions and (the real
$2k+1$th roots of) eigenvalues
\begin{equation}
\{\phi_{n}\}, \ \ \ \ \{a\lambda_{n}\}
\end{equation}
as for $H_{(2k+1)}$ in (A.10). Note that the operator 
$\Gamma_{5}=\gamma_{5}-H_{(2k+1)}=\gamma_{5}-H^{2k+1}$, which 
reverses the signature of eigenvalues of ``paired states'' and  
defines the index, is identical to (A.2) and (A.5).
 
We can confirm the second 
constraint  in (A.4) and the defining algebraic relation (A.2) 
for {\em any} $\phi_{n}$ in (A.17) by using (A.16),
\begin{eqnarray}
&&[H^{2}\gamma_{5}-\gamma_{5}H^{2}]\phi_{n}=0\nonumber\\
&&[\Gamma_{5}H+H\Gamma_{5}]\phi_{n}=0.
\end{eqnarray}

The general representation of the algebra (A.1) is obtained 
from the {\em standard representation}, which is defined by $H$ 
and $\{\phi_{n}\}$ in (A.17), and $\gamma_{5}$ in (A.16), by 
applying a suitable unitary transformation.

\end{document}